\documentclass[11pt,twoside]{article}
\usepackage{./asp2014}
\newcommand{\msun}{\mbox{M$_{\odot}$}}
\newcommand{\microjy}{\mbox{$\mu$Jy}}

\aspSuppressVolSlug
\resetcounters

\begin{document}

\title{New Frontiers in Protostellar Multiplicity with the ngVLA}
\author{John J. Tobin}
\affil{University of Oklahoma, \email{jjtobin@ou.edu}}
\author{Patrick Sheehan}
\affil{University of Oklahoma}
\author{Doug Johnstone}
\affil{NRC-Herzberg}

\paperauthor{John J. Tobin}{jjtobin@ou.edu}{}{University of Oklahoma}{}{Norman}{OK}{73072}{USA}
\paperauthor{Patrick D. Sheehan}{sheehan@ou.edu}{}{University of Oklahoma}{}{Norman}{OK}{73072}{USA}
\paperauthor{Doug Johnstone}{doug.johnstone@gmail.com}{}{NRC-Herzberg}{}{Victoria}{BC}{}{CA}

\begin{abstract}
The ngVLA will enable significant advances in our understanding of the formation and 
evolution of multiple star systems in the protostellar phase, building upon the
breakthroughs enabled by the VLA.
The high-sensitivity and resolution at 3~mm wavelengths and longer will enable closer 
multiple systems to be discovered in the nearby star forming regions. 
 The ngVLA is incredibly important
for multiplicity studies because dust opacity at short wavelengths ($<$3~mm) can hide multiplicity
and the long wavelengths are needed to reveal forming multiples in the youngest systems.
The samples sizes 
can be expanded to encompass star forming regions at distances of at least 1.5~kpc,
enabling statistical studies
that are on par with studies of field star multiplicity. We verify the capability of the 
ngVLA to detect and resolve multiple star systems at distances out to 1.5~kpc using empirical
examples of systems detected by the VLA and scaling them to greater distances. We also use
radiative transfer models and simulations to verify that the ngVLA can resolve close binary
systems from their dust emission at these distances. The ngVLA
will also have excellent imaging capability and the circum-multiple environments can also be examined 
in great detail.
\end{abstract}

\section{Introduction}

Star formation occurs as a consequence of dense gas clouds collapsing under their
own gravity, once the gravitational force is able to overcome 
sources of support \citep[e.g., thermal pressure, 
magnetic fields, turbulence;][]{mckeeostriker2007}. The star formation process frequently results
in the formation of two or more stars that comprise a gravitationally 
bound system, given that nearly half of Sun-like stars (in terms of 
stellar mass) are found in binary or higher-order multiple systems \citep{dm1991,raghavan2010}. The degree
of stellar multiplicity strongly depends on stellar mass. Stars more massive than the Sun have
a higher fraction of multiplicity and stars less massive than the 
Sun have a lower degree of multiplicity, but still with a multiplicity fraction upwards of 25\%;
see \citet{duchene2013} for a recent review. Thus, multiple star formation is a common
outcome of the star formation process for all stellar masses and a comprehensive understanding of star formation
must account for multiplicity.

Recent surveys in the infrared, millimeter, and centimeter  
have also shown a high degree of multiplicity in the protostar phase \citep{connelley2008,looney2000}.
Both the youngest protostars \citep[Class 0 sources,][]{andre1993} and more evolved protostars (Class I),
show a higher degree of multiplicity than field stars \citep[e.g.,][]{reipurth2004, chen2013,tobin2016a}.
The largest mass reservoir is available during the protostellar phase, making
it the most promising epoch for companion star formation to occur \citep{tohline2002}. Thus, the distribution
of field star multiplicity is likely derived from the primordial distribution of companions that form during
the protostar phase. While the mechanisms of multiple star formation are still uncertain, the
distribution of separations in the protostellar phase may reveal the signature of their 
formation process. The peak
of the companion separation distribution for field solar-type stars 
is $\sim$50 AU \citep{raghavan2010}, but
the formation route for these systems cannot be determined from evolved stellar populations
alone because they have undergone Myr to Gyr of dynamical evolution. 
Moreover, not all systems that form as multiples may remain multiple throughout
their lives, but they may still have formed in the presence of one or more companion stars
\citep{sadavoy2017}. 
Some companions can be ejected through dynamical interactions \citep[e.g.,][]{reipurth2012} 
or become unbound after the dispersal of the star forming core. This means that the formation of 
nascent planetary systems may have been influenced by companion stars, even if they are no longer
bound to the system. Thus, to reveal the origins of stellar multiplicity and its effects on 
proto-planetary systems, multiplicity must be characterized during the earliest stage 
of the star formation process.

There are two favored routes to explain the formation of multiple star systems: disk fragmentation
due to gravitational instability \citep[e.g.,][]{stamatellos2009, kratter2010} and 
turbulent fragmentation within the molecular cloud \citep[e.g.,][]{padoan2002,offner2010}. 
Disk fragmentation will preferentially result
in the formation of close ($<$500AU) multiple star systems and requires a large (R$_{disk}$ $\sim$50~AU)
and massive rotationally-supported disk to have formed around the primary star. 
Turbulent fragmentation can result in the formation of both wide and close 
multiple systems. In this scenario, the initial protostars form with separations
$\sim$1000~AU, and depending on their relative motions and masses they can migrate inward to
radii $\sim$100~AU \citep{offner2010}, remain at wide radii, or drift further apart. However, the expected trends can only
be revealed statistically, requiring large samples of protostars to be observed with spatial resolution
better than 50~AU (the average field star separation). It is also important to point out
that rotation of the protostellar cloud itself has also been suggested as a mechanism to form multiple
star systems \citep{bb1993}, but current observational evidence points more toward turbulence for
the formation of wide companions \citep{lee2016}.

A key difference in examining the formation of multiple stars at radio/millimeter wavelengths versus
optical/infrared is that direct emission from the protostellar photosphere is not being detected.
Instead, the dust emission from the individual circumstellar disks and/or free-free emission
from the base of the protostellar jet are being probed. Thus, the observations are tracing fragmentation, but
cannot directly confirm the protostellar nature of the multiple observed sources. 
\citet{tobin2016a} examined the nature of the emission detected and concluded that multiple sources of emission 
very likely correspond to true multiplicity, but we note, however, that 
the possibility of a small percentage of false positives cannot be excluded. Nevertheless,
observations at millimeter/centimeter wavelengths are generally the only available tool
to study multiplicity toward such deeply embedded objects.

The multiplicity statistics of field stars have the benefit of large samples 
and are not subject to the same limitations of protostellar multiplicity studies. 
Nonetheless, it is useful to use the field multiplicity studies
as a baseline for the survey requirements of future protostellar multiplicity studies.
The most recent compendium of field solar-type star multiplicity had a sample of 454 stars in
a volume limited sample, finding a total of 259 companions with an average companion separation
of 50~AU in an approximate Gaussian distribution \citep{raghavan2010}. 
Assuming that the field multiplicity distribution represents the distribution of protostellar multiplicity 
(it probably does not), we can use this distribution in order to estimate how many
protostellar multiples would need to be detected in order to obtain the same level of statistical
accuracy. Restricting the relevant parameter space to a separation range between 1~AU and 1000~AU,
the \citet{raghavan2010} sample contains $\sim$126 companion stars within this range. It is necessary
to limit the parameter space in this regard because scales less than 1~AU will likely be difficult
to examine for protostars and scales greater than 1000~AU will be dominated by clustering and
not physical association. In order to observe 126 protostellar companions to match the sample size,
at least 630 protostars must be observed if the average companion frequency of protostars is
in this range is 20\% \citep{tobin2016a}. This would require that all protostars within the 
Gould Belt be observed, sampling over different star formation conditions. If the youngest 
(Class 0) protostars are required for such characterization, then there are much less than 
630 in the entire Gould Belt \citep{dunham2014}. Surveys of more distant, massive star 
forming regions are therefore required to observe this number of protostars
in a single region.

\section{Setting the Stage: The VANDAM Survey}
A first large, systematic survey to examine protostellar multiplicity 
in an entire star forming region was carried out by the VLA Nascent Disk
and Multiplicity (VANDAM) Survey \citep{tobin2015}. This survey was carried 
out toward the Perseus star forming 
region, observing 45 Class 0 protostars (including possible objects in transition; Class 0/I), 
37 Class I protostars, and 12 Class II young
stars. All protostars were observed at a wavelength of 9~mm (33 GHz), at a 
uniform sensitivity of $\sim$9~$\mu$Jy, and at a uniform resolution
of 15~AU ( 0.065\arcsec). From these data, we identified 18 systems 
that were multiple with companion separations $<$500 AU; 16 of these 
companions were new detections by the VANDAM survey. We also detected a number of systems
with separations $>$500~AU as well as many hierarchical multiple systems. Some multiple systems had
separations as small as $\sim$19~AU, near the limit of our spatial resolution. 
Figure 1 shows an example of a very close
multiple system. Figure 2 then shows an example of a hierarchical system toward 
L1448 IRS3 where there is a triple, a binary,
and a single system all within 5000~AU. The triple was revealed by ALMA to have a circum-multiple disk, strongly indicating that the system formed via disk fragmentation \citep{tobin2016b}, and the 
system as a whole in Figure 2 illustrates how fragmentation is a multi-scale process.

\begin{figure}
\begin{center}
\includegraphics[scale=0.25]{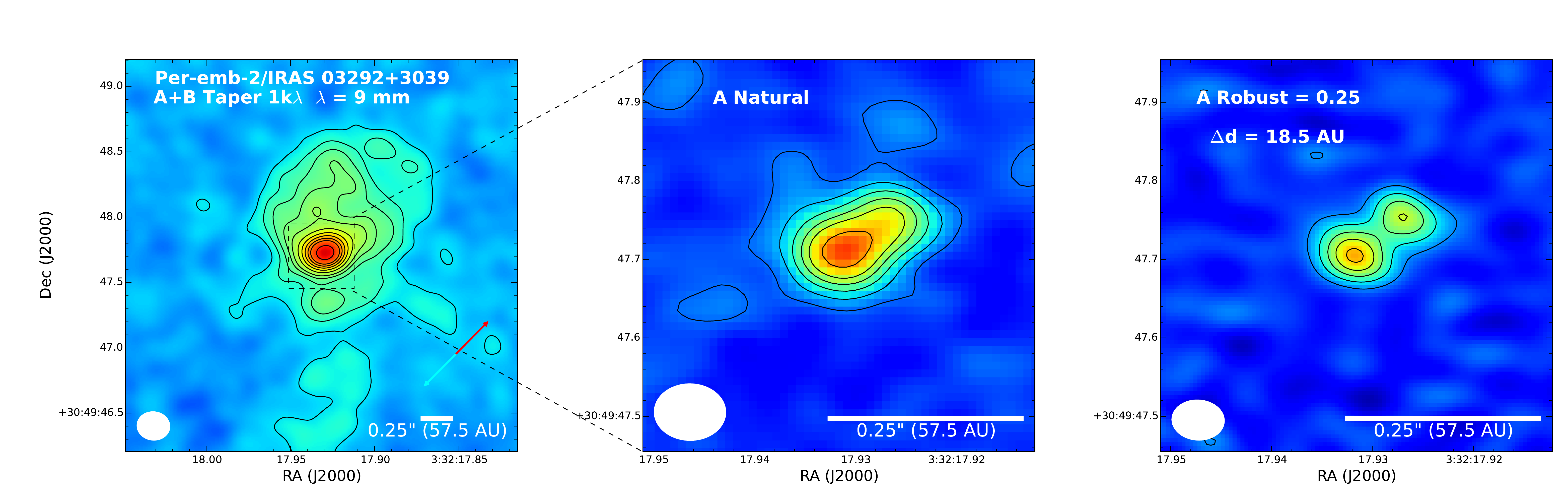}
\end{center}
\caption{Images of Per-emb-2 (IRAS 03292+3039) at 9~mm from the VLA
at increasing resolution from left
to right. The left panel, with the lowest resolution and most sensitivity 
to extended structure, shows significant/structured emission surrounding
a bright source that we interpret as the position of the main protostar; 
the middle and right panels zoom-in on the region outlined
with a dashed box. The middle panels with higher resolution have 
resolved-out the extended structure and only detect the bright peak at the position 
of the protostar; however, the source appears extended at this resolution.
The highest resolution image in the right panel shows that the
source is resolved into two sources separated by 18.5 AU.
The contours in each panel are [-6, -3, 3, 6, 9, 12, 15, 20, 25, 
30, 35, 40, 50, 60, 70, 80, 90, 100, 150] $\times$ $\sigma$, where
$\sigma$ = 7.3 \microjy, 9.6 \microjy, 11.9 \microjy\ from left 
to right at 9 mm.}
\end{figure}

\begin{figure}[!ht]
\begin{center}
\includegraphics[scale=0.8]{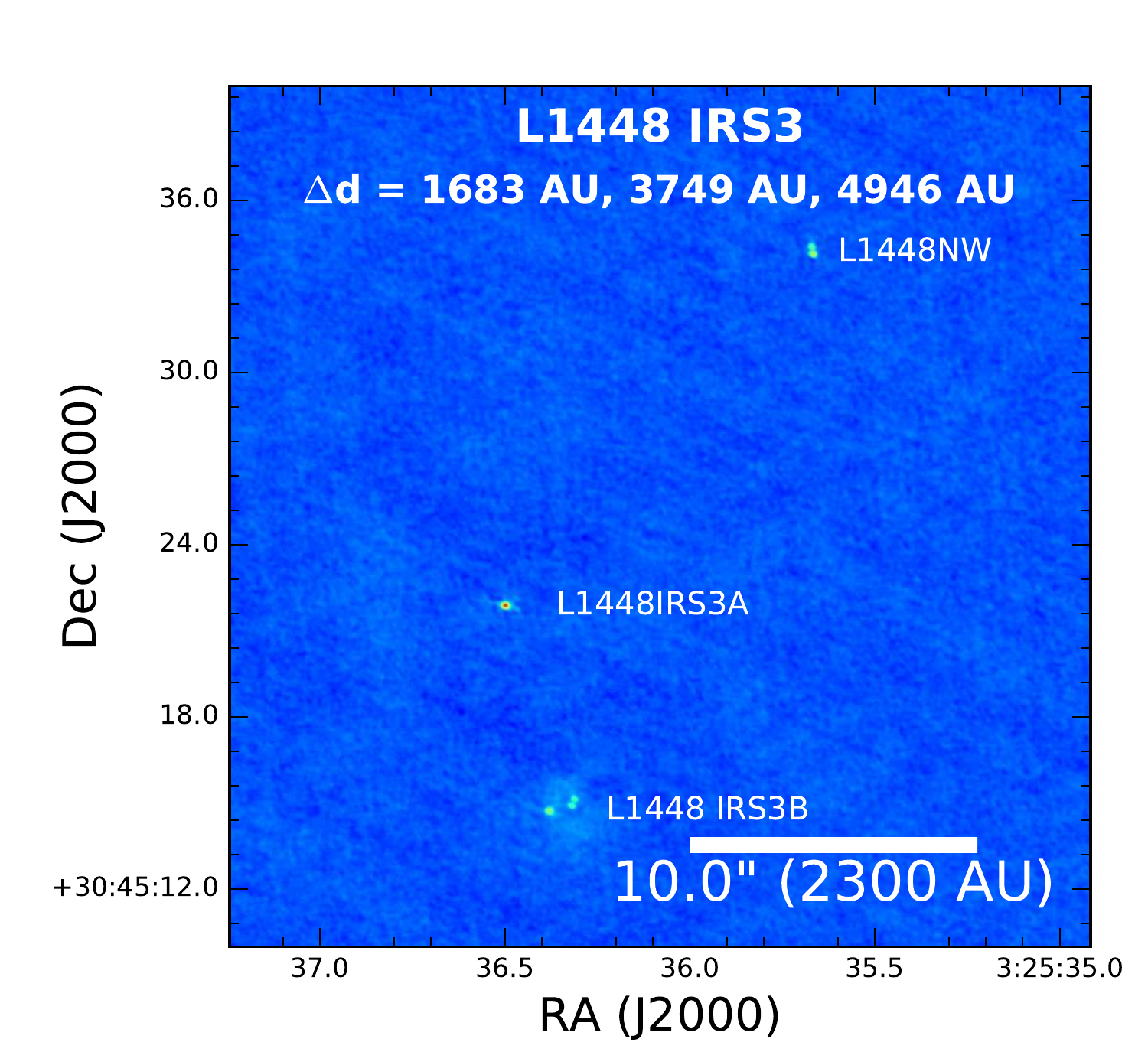}
\end{center}
\caption{Image of the L1448-N or IRS3 region at 9~mm from the VLA.
L1448 IRS3B is a close triple, L1448NW (L1448 IRS3C) is a binary, and L1448 IRS3A is 
single at the limit of our resolution. Both L1448 IRS3B and L1448NW are Class 0 sources 
while L1448 IRS3A is a Class I system.
Separations written inside the figure are relative to L1448 IRS3B.
}
\label{L1448N-wide}
\end{figure}

The separations of all companion stars detected in the VANDAM survey, are shown as a histogram
in Figure 3. Two features are obvious: 1) there appears to be two peaks, one at $\sim$75~AU and
another $>$1000~AU, and 2) the separation distribution
is in excess of the field, except for separations between $\sim$300~AU and $\sim$1000~AU.
It is argued in \citet{tobin2016a} that this bimodality results from both 
disk fragmentation ($\sim$75~AU part) and turbulent fragmentation 
($>$1000~AU part) happening to produce the observed distribution.
There are a lower number of detected companions between $\sim$300~AU and $\sim$1000~AU, 
indicating that neither formation mechanism is efficient at these scales.

While this survey of Perseus represents a major advance, it was 
limited in terms of the number of sources. Considering
uncertainties, the distribution could be consistent with a flat separation 
distribution (when considering the logarithm of the separation).
Thus, it is clear that greater numbers are required and more than one 
star forming region needs to be examined in order to
understand if this is a common distribution or an outlier. 
However, Perseus still offers an excellent template for
studies of multiplicity to larger samples and more distant star 
forming regions. It remains to be seen if the separation distribution observed 
toward Perseus is `universal,' and its implications for the formation mechanisms
of multiple stars are just beginning to be understood \citep{tobin2016b,lee2016,offner2016}.

Expanding on the discussion at the end of the Section 1, it will be important 
to determine if the observed bimodality is statistically 
significant. One way to do this is to obtain enough statistics such that
the uncertainties in the individual bins of Figure 3 are statistically 
distinct. Assuming the same distribution of companion separations in 
Perseus, the bin at 1.875 log(AU) will be inconsistent with the bin at 2.375 log(AU)
at the 4$\sigma$ level when the sample size reaches $\sim$500. Thus, 
samples in excess of 500 protostars are
needed to statistically distinguish between bimodal and log-flat distributions.

\begin{figure}
\begin{center}
\includegraphics[scale=0.65]{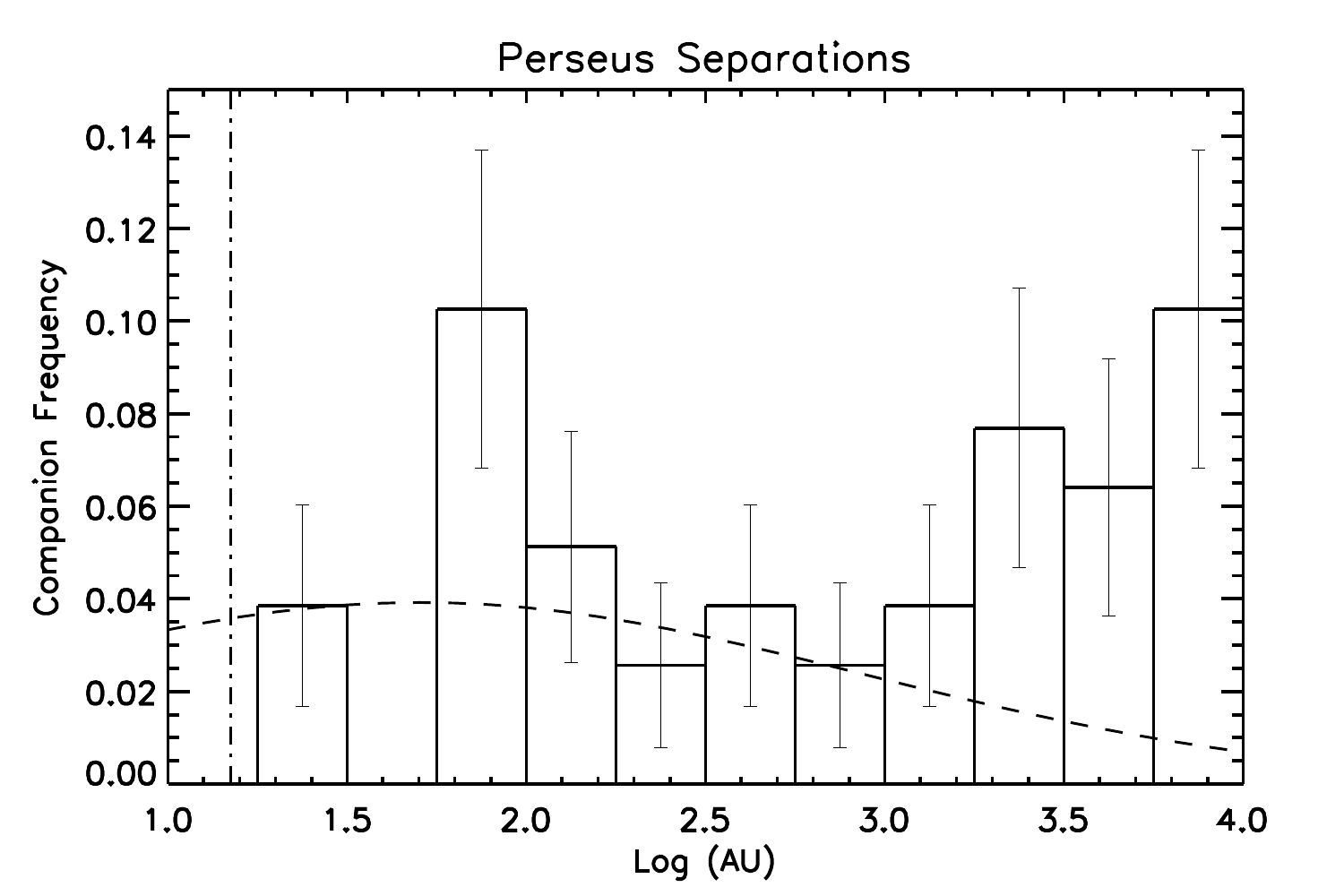}
\end{center}
\caption{Histogram of companion star frequency versus separation for the entire
sample of multiple sources in Perseus. The dashed
curve is the Gaussian fit to the field star separation distribution
from \citet{raghavan2010}. The vertical dot-dashed line 
corresponds to the approximate resolution limit of the VLA in A-configuration ($\sim$15 AU)
toward Perseus at a distance of 230~pc.}
\end{figure}

A challenge to observing larger numbers of protostars is that the most 
populous star forming regions are more distant. Orion is
$\sim$400~pc away and the VLA offers a best spatial resolution of
$\sim$30~AU at 8~mm, this is just a bit better that
the typical separation of field solar-type stars and there are likely to be 
more multiple systems lurking at separations below
30~AU. Perseus has at least three systems with separations $\sim$20~AU (Figures 1 and 3). 
Another limiting issue in Orion is that sources
are 3$\times$ fainter due to the increased distance, so observing 
with the same sensitivity requires 9$\times$ more observing time 
as compared to Perseus. It quickly becomes impractical to observe
even more distant star forming regions with sensitivity to protostars having
typical luminosities \citep[i.e., 1-3~L$_{\odot}$;][]{fischer2017}.

Finally, it is important to highlight that in the VANDAM survey, 
the spatial association of young stars was observed and not boundedness, which could not be evaluated
from the data at hand. Many (or most) of the companions shown in Figure 2 at separations $>$~1000~AU do
not likely reflect bound systems. Some could be line of sight associations (more frequent with increasing separation), or when they finish accreting material, outflows may have removed
enough mass such that they are unbound. The closer companions ($<$ 500~AU) have a higher likelihood of
being bound, but even so, interactions may alter or destroy some of these multiple systems. Thus,
the observed companion separation distribution in the protostellar phase may reflect the initial companion
separation distribution and it will dynamically evolve over Myr to Gyr toward toward the 
field star separation distribution.

The ngVLA has excellent potential to evaluate the boundedness of systems with $<$50 AU separations.
The angular resolution in the nearby regions will uncover closer multiple systems, and the positional accuracy
granted by the increased angular resolution will enable the orbital motion of $<$50~AU companions
to be measured on time scales of $\sim$5 years. Such analysis can incorporate previous VLA data
as needed to establish likely orbital solutions with an extended time baseline. Such measurements of
binary/multiple orbits will also offer constraints on the protostar masses.

\section{Examining Close Multiplicity Studies out to $\geq$1.5~kpc}

The ngVLA offers a number of features that overcome the limitations of the current VLA. The full array 
will provide an angular resolution of 0.01\arcsec\ at 7.3~mm and 
0.014\arcsec\ at 1.1~cm (ngVLA Memo \#17). These two wavelength
bands are ideal for examining protostellar multiplicity because they probe both dust and free-free emission. The addition of free-free emission to the dust emission
makes the protostars themselves stand out, enhancing their detectability. The 
dramatically increased resolution of the ngVLA will enable the range of separations 
to be expanded, with better sensitivity to close companions.
For Perseus (d=230~pc), we will be able to look for companions down to 2.3~AU separations. Toward Orion (d=400~pc), 
the separation limit will be pushed down to 4~AU, and in more distant, more populous star forming regions
(e.g., Cygnus-X at 1500~pc) the separation limit will be 15~AU. The ngVLA will enable us to examine separations 
in these distant Galactic star forming regions down to the same scale that we are currently able to do in Perseus!

Angular resolution improvements alone, however, are not sufficient to address questions of protostellar
multiplicity, because these studies require statistics and not a few case studies. With the drastic increase
in the number of antennas, and hence the collecting area, projected for the ngVLA, it is important to consider
the detectability of protostellar multiple systems in progressively more distant star forming regions.
Using IRAS 03292+3039 from Perseus (Figure 1) as a case study, current VLA observations detect 
sources A and B with peak flux densities of 170~$\mu$Jy~beam$^{-1}$ and 130~$\mu$Jy~beam$^{-1}$, respectively. If a similar
system were located at the distance of Cygnus-X, their respective flux densities would be
4~$\mu$Jy and 3~$\mu$Jy. These flux densities are completely impractical to observe at 30~GHz for the 
current VLA; reaching a S/N of 10 with the VLA would take 10 days! However, the ngVLA can reach a sensitivity of 0.26~$\mu$Jy~beam$^{-1}$ (S/N$\sim$10) in 
1~hour on-source. Thus, detecting such faint and close companions in massive, distant star forming
regions will be routine with the ngVLA because the increased sensitivity.

We have concentrated on the capabilities of the ngVLA at $\sim$7.3~mm because of the
ability to directly compare with data from the VANDAM surveys. However, the 3~mm band of the ngVLA 
also holds significant potential to enable studies of multiplicity out to star
forming regions even more distant than Cygnus-X. At 3~mm, most emission will be from pure dust emission, and
the brightness of this emission increases steeply with decreasing wavelength 
$\lambda^{-(2+\beta)}$ (increasing with frequency $\nu^{2+\beta}$), where $\beta$ is the dust opacity spectral
index which is typically observed to be between 0 and 2 in disks \citep[e.g.,][]{ricci2010}.
IRAS 03292+3039 was also observed with ALMA at 3~mm using 14~km baselines, obtaining nearly the highest resolution offered by the facility at 3~mm. The 3~mm flux densities of the two companions were
1200~$\mu$Jy and 1050~$\mu$Jy, respectively. Scaling these to the distance of Cygnus-X
they could be detected with S/N > 20 with the ngVLA in 1 hr on-source with a sensitivity of 
0.86~$\mu$Jy~beam$^{-1}$. At an even greater distance of $\sim$3~kpc, they could be detected
with a S/N=8 in 1 hr as well! Thus, the increased sensitivity of the ngVLA, coupled with the
higher angular resolution will radically enhance the study of multiple star formation. Furthermore,
the addition of a 3~mm band to the ngVLA would be particularly advantageous for the detection of 
multiple protostars in even more distant regions than possible with the 7.3~mm band due to the better
resolution and the fact that dust emission increases steeply with decreasing wavelength.

Obtaining sensitive images will still take a few hours of time with overheads
toward the more distant star forming regions, but star forming regions at greater distances
occupy less area of the sky in terms of their solid angle. The wide field of view offered 
by the ngVLA (a factor of 2 increase in solid angle over the VLA), coupled with the 
increased sensitivity, means that many more protostars can be captured in a single pointing
than in the nearby star forming regions, this will somewhat offset the need for longer 
integrations times because more sources will be observed in a single observation.

While ALMA of course has unmatched sensitivity to dust emission at shorter wavelengths, the
shorter wavelength becomes a disadvantage when attempting to study the youngest protostars.
On scales less than $\sim$100~AU, the dense inner envelope or disks of protostars can 
be opaque at wavelengths as long as 1.3~mm and 3~mm, hiding small-scale multiplicity (see
Chapter `Exploring Protostellar Disk Formation with the ngVLA'). 
Moreover, at higher frequencies
the field of view of ALMA becomes small, making it unable to simultaneously observe as
many protostars at high resolution. Thus the ngVLA capability of
0.01\arcsec\ or better resolution at wavelengths between 2~cm and 3~mm is an absolutely
unique and critical capability to examine the formation of multiple stars during the
early stages of protostellar evolution.

Because detection of dust emission from the small circumstellar disks in multiple
systems is the most well-defined route to detection, it is important to further demonstrate
their detectability. We ran a few radiative transfer models of protostars with small
disks in binary star systems and simulated their observation with the full ngVLA
using the CASA \textit{simobserve} task. The modeling is described in more detail
in the Chapter `Exploring Protostellar Disk Formation with the ngVLA.' We show
the results of our simulations in Figure 4, where we have simulated the observation
of two protostars, separated by 5~AU, each with a 1~AU radius and 0.001~\msun\ disk. These
protostars are simulated at a distance of 400~pc, and they can be well-detected by
the ngVLA with S/N = 20 in 1 hr. We also computed a model for a 15~AU separation binary system
at a distance of 1.5~kpc and each disk having a radius of 3~AU and a mass of 0.01~\msun. 
The 1.5~kpc binary can be detected by the ngVLA with S/N = 10 in 1 hr. Thus,
the ngVLA will enable the detection of extremely close binary protostars 
with separations as small as 3~AU (at 230~pc) from their dust emission alone, 
under the assumption that circumstellar disks have radii of order
their separation/3 \citep{artymowicz1994}. 
However, if each component has some free-free emission, in addition 
to the dust, they can likely be resolved at closer separations and their disks
would not need to be as massive.

In addition to the ngVLA's ability to examine extremely 
close companions, the imaging capabilities of the ngVLA are superb enough 
to enable circum-multiple environments to be characterized. Using an ALMA image at 0.87~mm of L1448 IRS3B system,
a triple system with a surrounding circum-multiple disk with spiral structure \citep{tobin2016b}, we scaled
the surface brightness assuming optically thin dust emission and $\beta$=1, implying a flux density
scaling as $\lambda^{-3}$ (see discussion in preceding paragraphs). Then we simulated a 1 hour observation
at 3~mm, and a 2 hour observation at 7.3~mm using the 168 antenna `Plains array,' keeping 
the distance at 230~pc. We show the results of modeling 
in Figure \ref{IRS3B}; the structure observed at 0.87~mm is
fully recovered with high fidelity at 3~mm and the image is also well-recovered at 7.3~mm
but with lower S/N due to the fainter dust emission. Thus, the ngVLA will also enable the imaging
of circum-multiple environments simultaneously while probing for multiplicity. At 3~mm
and 7.3~mm, the sensitivity of the ngVLA will superior to that of ALMA at these wavelengths
and have higher angular resolution even with only the `Plains' array.
The circum-multiple emission is more likely to be optically thin at these wavelengths as
compared to shorter wavelengths with ALMA.

The ngVLA will open up three exciting regions of parameter space in protostellar multiplicity
studies. It will conduct high-S/N imaging toward nearby star forming regions (i.e., Perseus, Taurus)
with enough angular resolution to search for companion protostars that have $\sim$5$\times$ smaller 
separation than can be 
examined with the VLA. The ngVLA will also enable studies of protostellar multiplicity
to obtain far greater statistics through the simultaneous observation of larger
numbers of protostars in more distant regions (e.g., Cygnus-X) with the same or better sensitivity than 
is currently possible toward the nearby regions. Finally, the ngVLA can examine the environments
around multiple star systems with high-S/N, with best results at the shortest wavelengths.

\begin{figure}
\begin{center}
\includegraphics[scale=0.65]{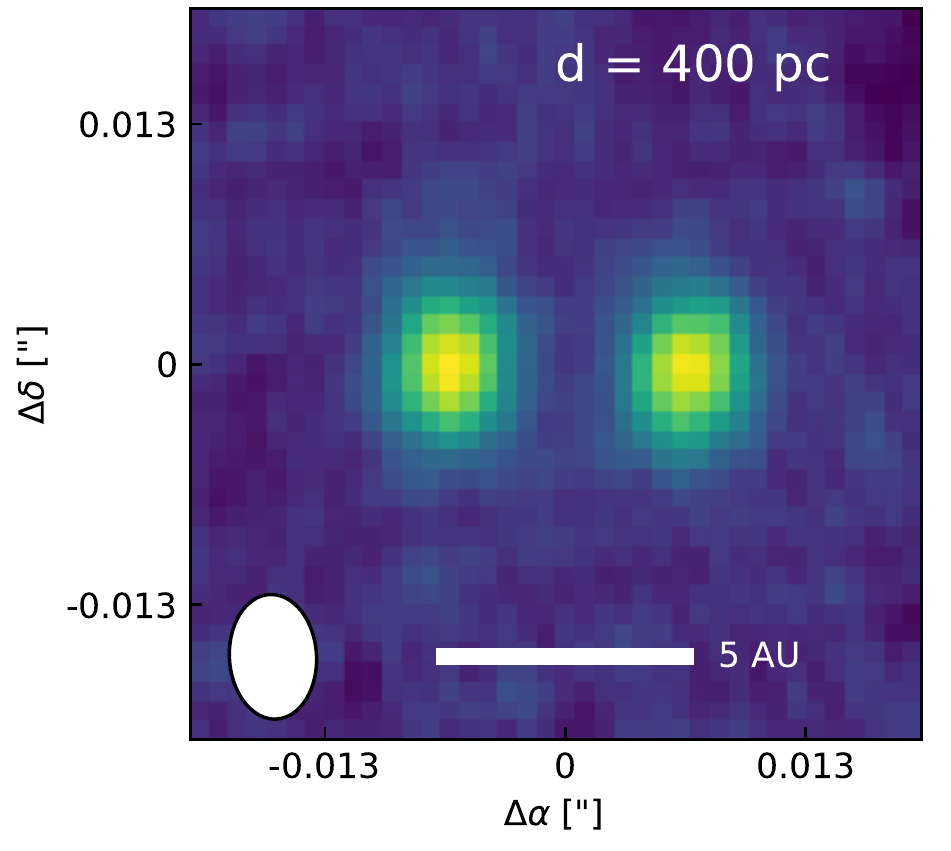}
\includegraphics[scale=0.65]{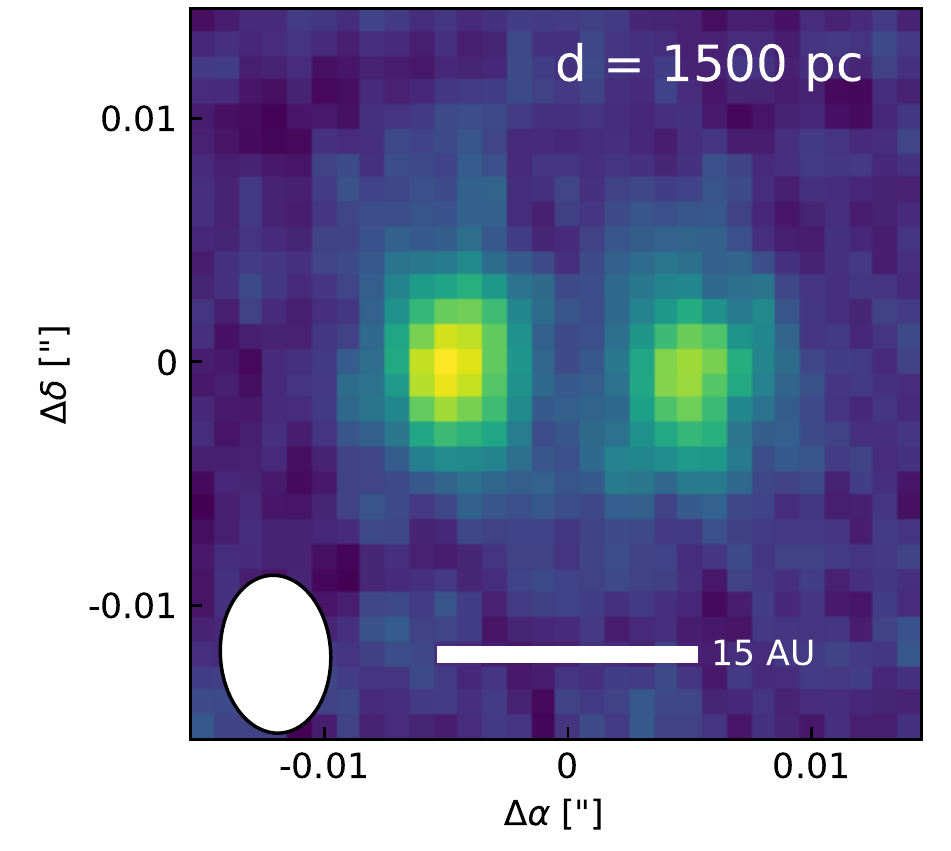}
\end{center}
\caption{Synthetic ngVLA observations of binary systems comprised
of emission from compact circumstellar disks. The left panel is at
a distance of 400~pc having R$_{disk}$~=~1~AU and M$_{disk}$~=~0.001~\msun,
and the right pane is at a distance of 1.5~kpc R$_{disk}$~=~3~AU and M$_{disk}$~=~0.01~\msun. 
Thus, these models demonstrate the feasibility
of detecting small disks in multiple star systems at close separations.
}
\end{figure}

\begin{figure}
\begin{center}
\includegraphics[scale=0.275]{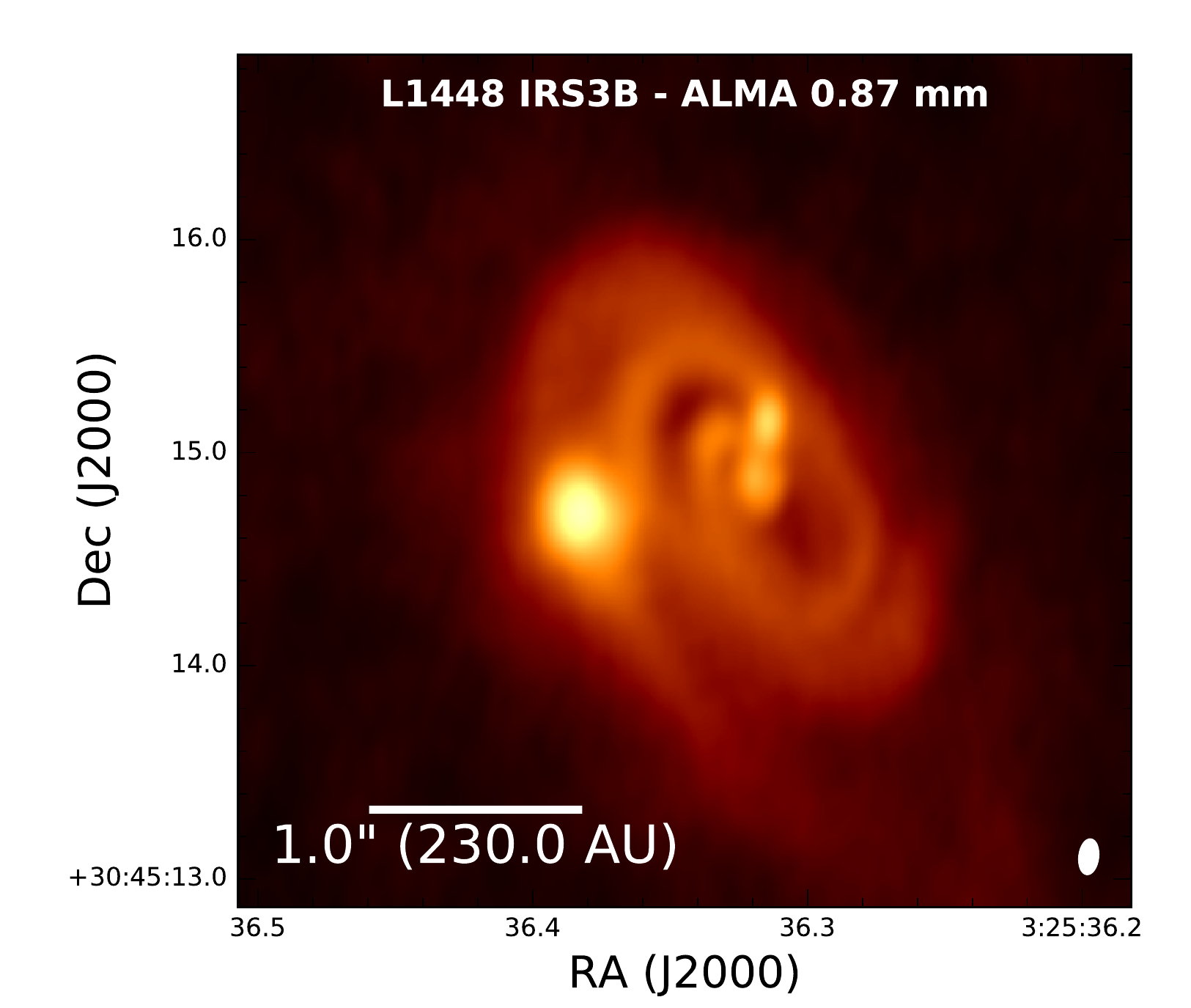}
\includegraphics[scale=0.275]{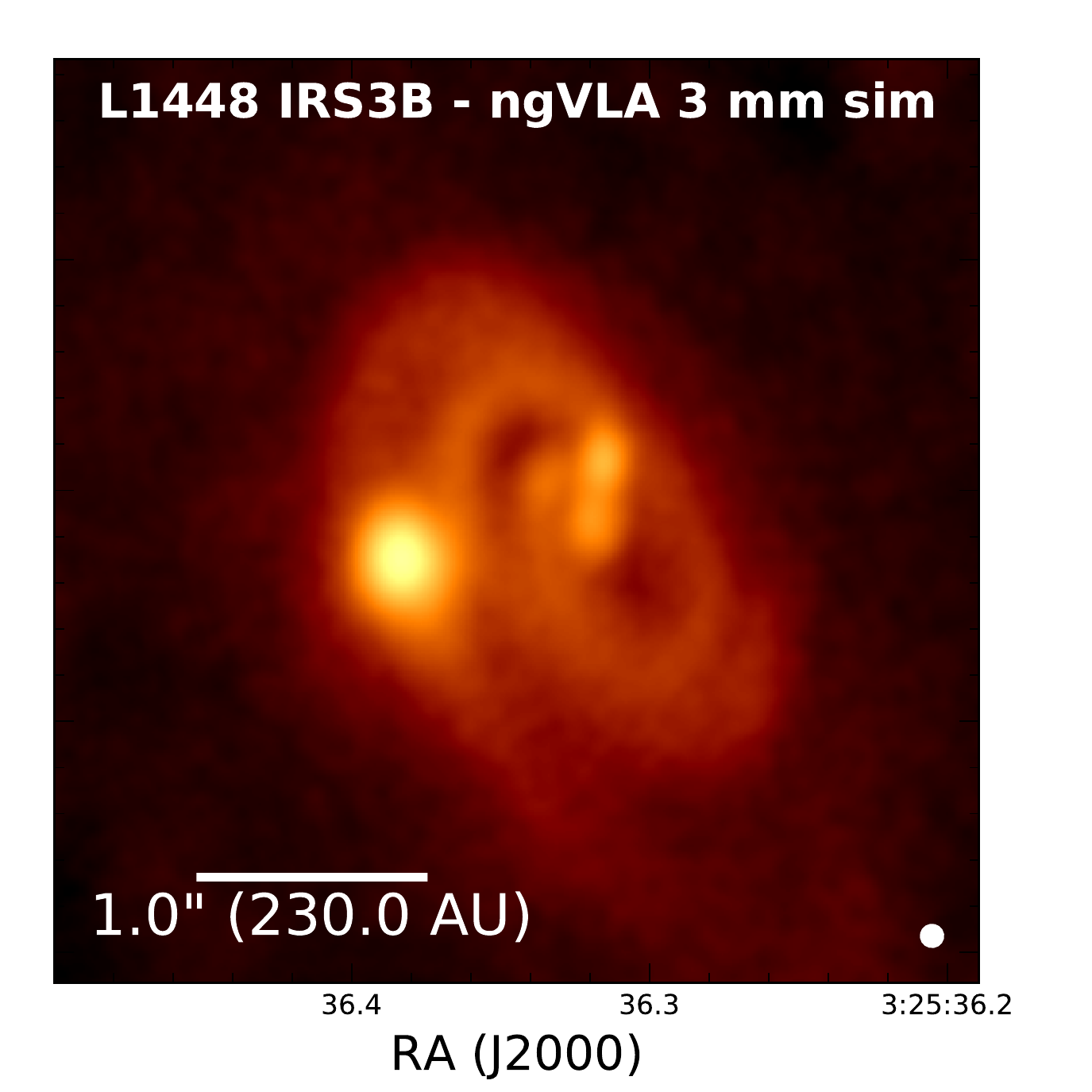}
\includegraphics[scale=0.275]{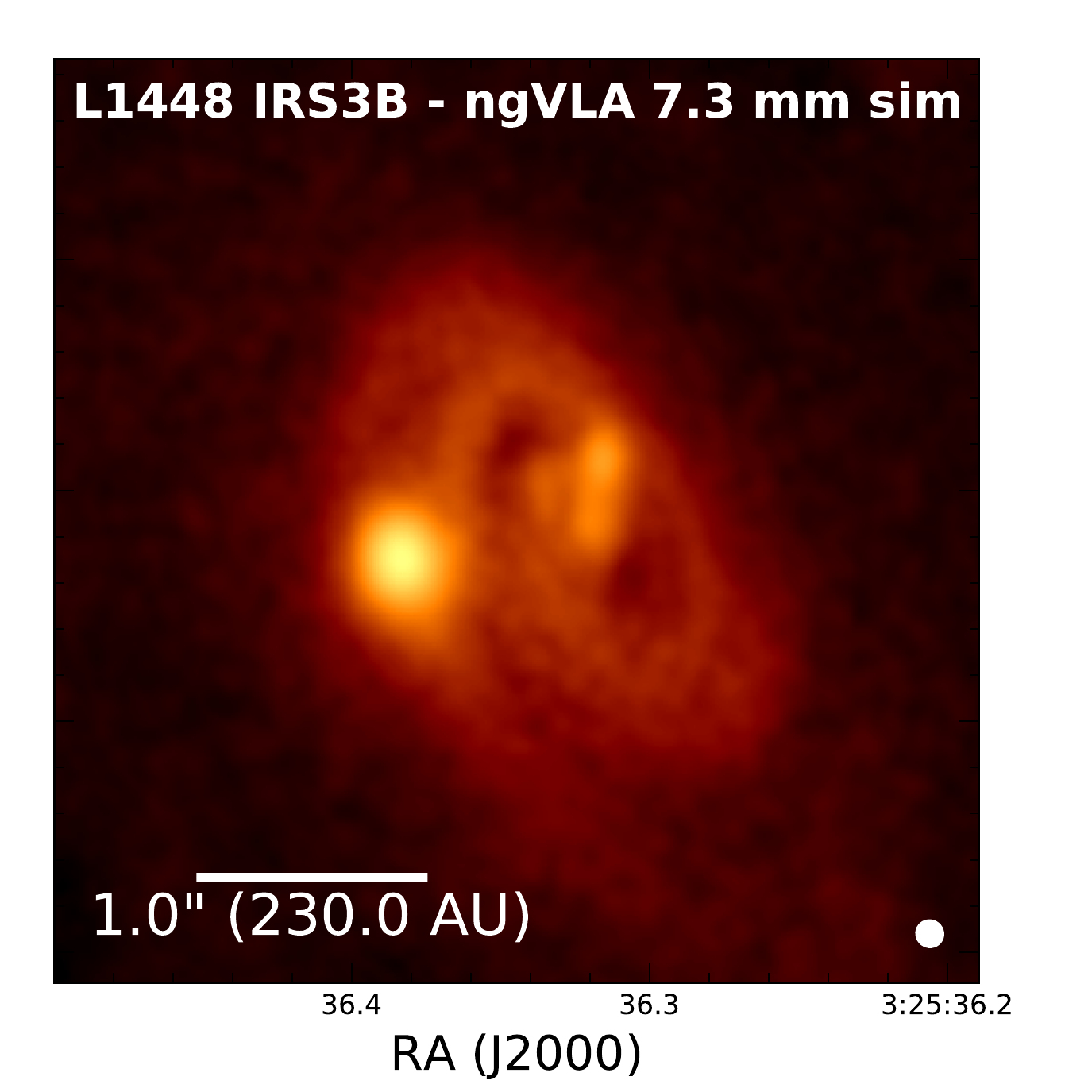}
\end{center}
\caption{Synthetic ngVLA observations of the triple system L1448 IRS3B using
only the 168 antenna `Plains' array. These simulations
used a scaled ALMA 0.87~mm image as the model (left) and utilize the maximum bandwidth available
for the respective ngVLA bands, 40~GHz (3~mm; middle) and 20 GHz (7.3~mm; right). The 3~mm simulation is one hour
on-source and is imaged using Briggs weighting, robust=0.5, and a taper at 2000~k$\lambda$;
the 7.3~mm simulation is two hours on-source and the rest of the imaging parameters are the same
as the 3~mm.
}
\label{IRS3B}
\end{figure}

\section{Synergies with Other Facilities}

While the ngVLA will be able to identify very close multiple systems perhaps out to 3~kpc,
ALMA will be very useful in characterizing the immediate surroundings of multiple star systems.
The brighter dust emission at shorter wavelengths may make ALMA an ideal complement to very high
resolution studies with the ngVLA, both in the dust continuum and molecular lines,
provided that the dust emission is not too optically thick. ALMA
will be able to provide access to the necessary molecular lines to characterize the immediate
environs around proto-multiple systems. This is because many of the most abundant, disk-tracing molecules
have their strongest rotation transitions at wavelengths shorter than 3~mm; the
ngVLA will only be able to access the ($J=1\rightarrow0$) transitions for most 
molecules, and these transitions do not generally have the brightest 
emission in the warm ($>$20~K) regions in the immediate vicinity of protostars. Furthermore,
ALMA dust continuum observations can detect circum-multiple disks around the protostar that 
might have column densities too low to detect with the ngVLA, but as shown 
in Figure \ref{IRS3B} the ngVLA will have this capability as well.

The more distant star forming regions have a disadvantage relative to the nearby ones in that their
protostellar content is not as well-characterized due to the low resolution ($>$1\arcsec) 
of previous mid-to-far-infrared
surveys. The James Webb Space Telescope will undoubtedly survey numerous massive star forming regions and 
infrared dark clouds (IRDCs) that harbor significantly more young stars than the nearby regions at wavelengths between 10 and 28~$\mu$m. Thus, by the time the ngVLA is conducting early science, 
the protostellar content of more distant star forming regions 
is likely to be much more well-characterized, enabling the multiplicity results obtained by the
ngVLA to be put into a similar context as the results toward nearby star forming regions.

\section{Summary}

The ngVLA will open a new window into the study of multiplicity during the protostellar
phase, providing a much clearer picture of where most companion stars are forming.
Thus, the ngVLA will improve our understanding of just how many stars and proto-planetary
disks begin their lives initially as part of a multiple system and how multiplicity affects
the evolution of planetary systems, whether or not the system remains a multiple in
its main sequence life. The increased resolution of the ngVLA will enable both
closer multiples to be detected than previously possible for protostellar systems, and
coupled with the increased sensitivity, larger numbers of multiple star systems can be
observed in order to obtain statistics that equal or surpass that of the field solar-type stars.
\bibliography{tobin_multiplicity}  



\end{document}